\documentstyle[12pt,latexsym]{article}
\begin{document}

\title{Multistability in dynamical systems\thanks{%
to appear in \textit{Dynamical Systems: From Crystal to Chaos}, World
Scientific, 1999}}
\author{R. Vilela Mendes \\
%EndAName
Grupo de F\'{i}sica-Matem\'{a}tica, \\
Complexo Interdisciplinar, Universidade de Lisboa,\\
Av. Gama Pinto 2, 1699 Lisboa Codex, Portugal}
\date{}
\maketitle

\begin{abstract}
In neuroscience, optics and condensed matter there is ample physical
evidence for multistable dynamical systems, that is, systems with a large
number of attractors. The known mathematical mechanisms that lead to
multiple attractors are homoclinic tangencies and stabilization, by small
perturbations or by coupling, of systems possessing a large number of
unstable invariant sets. A short review of the existent results is
presented, as well as two new results concerning the existence of a large
number of stable periodic orbits in a perturbed marginally stable
dissipative map and an infinite number of such orbits in two coupled
quadratic maps working on the Feigenbaum accumulation point.
\end{abstract}

\section{Multistability. Motivations.}

That are three basic motivations to study multistable systems, that is
systems that possess a large number of coexisting attractors for a fixed set
of parameters. First there is ample evidence for such phenomena in the
natural sciences, with examples coming, among others, from neurosciences and
neural dynamics \cite{Schiff} \cite{Foss} \cite{Seung} \cite{Freeman} \cite
{Dente} \cite{MacKay} \cite{Bohr} , optics \cite{Hammel} \cite{Brambilla},
chemistry \cite{Marmillot} \cite{Laplante} \cite{Hunt}, condensed matter 
\cite{Prengel} and geophysics \cite{Yoden}.

The second motivation to study multistable dynamical systems is the
mathematical challenge of identifying the universal mechanisms that lead to
multistability and to prove rigorously under what circumstances the
phenomenon ,may occur.

The third and last motivation arises from the field of control and
technological design. After the pioneering work of Ott,
Grebogi and Yorke \cite{Ott}, the field of control of chaos became a whole
industry (\cite{Control} and references therein). Control of chaos deals
with the local control of unstable orbits, either to achieve their
stabilization or, alternatively, to target a dynamical system to some
desired final state. Reliable stabilization of unstable periodic orbit
requires either the knowledge of a good model of the system or an accurate
local reconstruction of the dynamics. This is feasible in some
low-dimensional systems, but it seems rather problematic for
high-dimensional ones. The typical situation in control of chaos, is that of
a strange attractor with an infinite number of embedded periodic orbits, all
of them unstable. If, instead of an infinite number of unstable periodic
orbits, one has, for example, an infinite number of sinks, the controlling
situation would seem more promising. The sinks would of course have very
small basins of attraction. Nevertheless, the control need not be so
accurate, because it suffices to keep the system inside the desired basin of
attraction. This, in principle, makes for a more robust control.

In this paper I will concentrate mostly on the rigorous mathematical results
that concern multistable dynamical systems. Sects. 2 and 3 contain brief
reviews of the Newhouse phenomenon and of the creation of a large number of
periodic orbits by dissipative perturbations of conservative systems.
Finally, in Sects. 4 and 5, some new results are presented which, in Sect 3,
use the techniques of deformation stability and, in Sect. 5, prove the
existence of an infinite number of sinks for two coupled quadratic maps.

\section{Diffeomorphisms with homoclinic tangencies.The Newhouse phenomenon
and beyond}

Contrary to earlier conjectures that generic systems might have only
finitely many attractors, Newhouse \cite{Newhouse1} \cite{Newhouse2} \cite
{Newhouse3} proved that a class of diffeomorphisms in a two-dimensional
manifold has infinitely many attracting periodic orbits (sinks), a result
that was later extended to higher dimensions\cite{Palis}. Concretely, for
two-dimensional manifolds the result is:

\textbf{Theorem }(Newhouse, Robinson\cite{Robinson}) Let $f_{\mu }$ be a $%
C^{3}$ map in a 2-dimensional manifold with $C^{1}$ dependence on $\mu $ and 
$\left| \det \left( T_{a}f_{\mu _{0}}^{n}\right) \right| <1$ and let the
non-degenerate homoclinic tangency be crossed at non-zero speed at $\mu =\mu
_{0}$. Then for $\forall \varepsilon >0$, $\exists (\mu _{1},\mu
_{2})\subset (\mu _{0},\mu _{0}+\varepsilon )$ and a residual subset $%
J\subset (\mu _{1},\mu _{2})$ such that for $\mu \in J$ , $f_{\mu }$ has
infinitely many sinks.

Models of such diffeomorphisms were constructed by Gambaudo and Tresser\cite
{Gambaudo} and Wang proved that the Newhouse set has positive Hausdorff
measure\cite{Wang}.

After these results, intense research followed on the unfolding of
homoclinic tangencies and an essential question was whether, in addition to
infinitely many sinks, there would also be infinitely many strange
attractors near the homoclinic tangencies. The question was positively
answered by Colli\cite{Colli}. The main result is:

\textbf{Theorem }(Colli) Let $f_{0}\in Diff^{\infty }(M)$ be such that $%
f_{0} $ has a homoclinic tangency between the stable and unstable manifolds
of a dissipative hyperbolic saddle $p_{0}$. Then, there is an open set $%
\Omega \subset Diff^{\infty }(M)$ such that

(a) $f_{0}\in \overline{\Omega }$

(b) there is a dense subset $D\subset \Omega $ such that for all $f\in D$ , $%
f$ exhibits infinitely many coexisting H\'{e}non-like strange attractors.

Having established the existence of infinitely many sinks and infinitely
many strange attractors near homoclinic tangencies, a question of practical
importance is the stability of the phenomenon under small random
perturbations of the deterministic dynamics. It turns out that the answer to
this question is negative. Therefore under small random perturbations only
finitely many physical measures will remain.

\textbf{Theorem }(Ara\'{u}jo\cite{Araujo}) Let $f:M\rightarrow M$ be a
diffeomorphism of class $C^{r},r>1$, of a compact connected boundaryless
manifold $M$ of finite dimension. If $f=f_{a}$ is a member of a parametric
family under parametric noise of level $\varepsilon >0$, that satisfies the
hypothesis:

There are $K\in N$ and $\xi _{0}>0$ such that, for all $k\geq K$ and $x\in M$

(A) $f^{k}(x,\Delta )\supset B^{k}(x),\xi _{0})$ ;

(B) $f^{k}(x,\nu ^{\infty })<<m$ ;

then there is a finite number of probability measures $\mu _{1},\cdots \mu
_{l}$ in $M$ with the properties

1. $\mu _{1},\cdots \mu _{l}$ are physical absolutely continuous probability
measures;

2. supp$\mu _{i}\cap $supp$\mu _{j}$ for all $1\leq i<j\leq l$ ;

3. for all $x\in M$ there are open sets $V_{1}=V_{1}(x),\cdots
,V_{l}=V_{l}(x)\subset \Delta $ such that

(a) $V_{i}\cap V_{j}=\emptyset $, $1\leq i<j\leq l$ ;

(b) $\nu ^{\infty }\left( \Delta \backslash \left( V_{1}\cup \cdots \cup
V_{l}\right) \right) =0$ ;

(c) for all $1\leq i\leq l$ and $\nu ^{\infty }-$ a.e. $t\in V_{i}$ we have 
\[
\lim_{n\rightarrow \infty }\frac{1}{n}\sum_{j=0}^{n-1}\phi \left(
f^{j}(x,t)\right) =\int \phi d\mu 
\]
for every $\phi \in C\left( M,R\right) $. Moreover the sets $V_{1}(x),\cdots
,V_{l}(x)$ depend continuously on $x\in M$ with respect to the distance $%
d_{\nu }\left( A,B\right) =\nu ^{\infty }\left( A\triangle B\right) $
between $\nu ^{\infty }-\textnormal{mod}0$ subsets of $\Delta $.

\section{Small dissipative perturbations of conservative systems}

Conservative systems have a large number of coexisting invariant sets,
namely periodic orbits, invariant tori and cantori. By adding a small amount
of dissipation to a conservative system one finds that some of the invariant
sets become attractors. Of course, not all invariant sets of the
conservative system will survive when the dissipation is added. However, for
sufficiently small dissipation many attractors (mainly periodic orbits) have
been observed in typical systems. Poon, Grebogi, Feudel, Hunt and Yorke \cite
{Poon} \cite{Feudel1} \cite{Feudel2} have extensively studied these effects
in the single and double rotor, the H\'{e}non map and the optical cavity
map. They find a large number of attractors for a small amount of
dissipation, in particular in the double rotor map. The large number of
coexisting stable periodic orbits has a complex interwoven basin of
attraction structure, with the basin boundaries permeating most of the state
space. The chaotic component of the dynamics is in the chaotic saddles
embedded in the basin boundary. The systems are also found to be highly
sensitive to small amounts of noise. The authors have argued that the two
attributes, namely, accessibility to many different states and high
sensitivity, are an asset in the sense that they are suitable for an easy
control of the complex system. The transition between different stable
states, poses however delicate problems in view of the large chaotic
transients in the basin boundary. The problem of migration between
attractors and their stability in multiple-attractor systems has also been
studied by other authors\cite{Weigel} \cite{Kaneko}.

All this work is very interesting, however most of results are based on
numerical evidence. It would be desirable to have some control of the
effects by rigorous mathematical methods. The techniques of deformation
stability, to be discussed in the next section, might provide such a tool,
in some cases at least.

\section{Deformation stability}

The basic idea is that, when dissipation is added to a conservative system,
it is only a part of the total phase space that is related to the invariant
sets of the dissipative system. The question therefore is to find the
subsets in the conservative phase space that correspond to the stable sets
in the dissipative system. In a sense the situation is similar to KAM\
theory and to the role played by the constants of motion in this theory. In
KAM theory, the first integrals $I_{0}$ defined everywhere in phase space
for the integrable system, are deformed into a set of constants of motion $%
I_{\varepsilon }$ which are defined only over a subset of sufficiently
irrational tori. By analogy, a similar deformation stability may exist for
dissipative perturbations of conservative systems, the stable domain of the
constants of motion being, for example, the closure of a family of periodic
orbits. In the papers \cite{Vilela1} \cite{Vilela2} \cite{Lima} \cite
{Vilela3} where these ideas were developed, we were concerned with
permanence under deformation of invariant sets, not necessarily attracting
invariant sets. Most results may however be adapted to the search for
attractors. As an example a result will be proved which gives a rigorous
criterium for the existence of stable periodic orbits in a class of maps.

Let an $\varepsilon -$family of maps be 
\[
\begin{array}{lll}
x^{^{\prime }} & = & bx+y+f(x,y,\varepsilon ) \\ 
y^{^{\prime }} & = & y+g(x,y,\varepsilon )
\end{array}
\]
with $f(x,y,0)=g(x,y,0)=0$ For $\varepsilon =0$ the map has marginally
stable periodic orbits of all periods. Under perturbation some of the orbits
become stable ones.

\textbf{Theorem }If $f$ and $g$ are jointly $C^{2}$ in $(x,y,\varepsilon )$
with $f(x,y,0)=g(x,y,0)=0$, there is an $\overline{\varepsilon }$ such that
for $\left| \varepsilon \right| <\left| \overline{\varepsilon }\right| $ an
interior orbit of period $p$ of the unperturbed map becomes a stable orbit
of the perturbed map if and only if:

(1) $\sum\limits_{n=0}^{p-1}\partial _{\varepsilon
}g(x_{n}^{(0)},y^{(0)},0)\mid _{\varepsilon =0}=0$

(2) 
\[
\varepsilon \partial _{\varepsilon }\sum\limits_{n=0}^{p-1}\left\{ \partial
_{x}g(x_{n}^{(0)},y^{(0)},\varepsilon )+(1-b)\partial
_{y}g(x_{n}^{(0)},y^{(0)},\varepsilon )\right\} \mid _{\varepsilon =0}<0 
\]

\textbf{Proof}:

Iterating the map $p$ times the orbit condition is: 
\begin{equation}
\begin{array}{lll}
(b^{p}-1)x+y\frac{1-b^{p}}{1-b}+u_{p}(x,y,\varepsilon )-\beta _{p}(b) & = & 0
\\ 
v_{p}(x,y,\varepsilon )=0 &  & 
\end{array}
\label{4.1}
\end{equation}
with 
\begin{equation}
\begin{array}{lll}
u_{p}(x,y,\varepsilon ) & = & \sum\limits_{n=0}^{p-1}b^{p-k-1}f(x_{n},y_{n},%
\varepsilon ) \\ 
&  & +(1-b)^{-1}\sum\limits_{n=0}^{p-2}(1-b^{p-k-1})g(x_{n},y_{n},%
\varepsilon ) \\ 
v_{p}(x,y,\varepsilon ) & = & \sum\limits_{n=0}^{p-1}g(x_{n},y_{n},%
\varepsilon )
\end{array}
\label{4.2}
\end{equation}
and $\beta _{p}(b)$ is a polynomial in $b$.

$b<1$ and the $C^{2}$ condition imply, by the implicit function theorem, the
existence of a solution $x=h(y,\varepsilon )$ of the first equation in (\ref
{4.1}). The second equation becomes 
\begin{equation}
v_{p}(h(y,\varepsilon ),y,\varepsilon )=\varepsilon \overline{v_{p}}%
(x,y,\varepsilon )=0  \label{4.3}
\end{equation}
$\overline{v_{p}}$ being a $C^{1}$ function. Condition (1) in the theorem
follows from Eq.(\ref{4.3}) for $\varepsilon =0$ and the non-vanishing of
the expression in the condition is required by the application of the
implicit function theorem to (\ref{4.3}).

The eigenvalues of the $p$-iterated unperturbed map are $1$ and $b^{p}$. The
sign in condition (2) is required to perturb the first eigenvalue to a
smaller value. $\Box $

\section{Coupled quadratic maps}

A system as simple as one composed of two coupled quadratic maps, may have
an infinite number of stable periodic orbits. Let the system be

\begin{equation}
\begin{array}{c}
x_{1}(t+1)=1-\mu _{*}\left( (1-c)x_{1}(t)+cx_{2}(t)\right) ^{2} \\ 
x_{2}(t+1)=1-\mu _{*}\left( cx_{1}(t)+(1-c)x_{2}(t)\right) ^{2}
\end{array}
\label{5.1}
\end{equation}
with $x\in [-1,1]$, and $\mu _{*}=1.401155...$ , which is the parameter
value of the period doubling accumulation point.

\textbf{Theorem} \cite{Carvalho} For sufficiently small $c$\ there is an $N$%
\ such that the system (\ref{5.1}) has stable periodic orbits of all periods 
$2^{n}$\ for $n>N$.

\textbf{Proof}:

Two essential features in the proof are the permanence of the unstabilized
orbits in a flip bifurcation and the contraction effect introduced by the
convex coupling. Only a sketch of the proof will be presented. For more
details refer to Ref.\cite{Carvalho}.

The bifurcations leading to the Feigenbaum accumulation point at $\mu _{*}$
are flip bifurcations. This means that, after each bifurcation, the orbit
that looses stability remains as an unstable periodic orbit. Therefore, (for 
$c=0$) at $\mu =\mu _{*}$ the system (\ref{5.1}) has an infinite number of
unstable periodic orbits of all periods $p=2^{n}$.

The proof has two basic steps. First one proves that, for sufficiently small 
$c\neq 0$, these periodic orbits still exist in the system (\ref{5.1}).
Second, that for any such $c$, there is an $N$ such that there is at least
one stable orbit for all periods $p=2^{n}$ with $n>N$. For both steps an
important role is played by the instability factor, given by $\left(
f^{(p)}\right) ^{^{\prime }}(x_{p})$ at the fixed points $x_{p}$ of the $p-$%
iterated map. Using the properties of the Feigenbaum - Cvitanovic functional
equation one finds that the instability factor $\left( f^{(p)}\right)
^{^{\prime }}(x^{*})$ converges to a fixed non-zero uniformly bounded value
for all orbits. One now proceeds to the proof of the theorem.

\textit{First step}: Permanence of the periodic orbits for small $c$

Let $x_{p}^{*}\in [-1,1]\times [-1,1]$ be, for example, the coordinate of
the $p-$periodic orbit closest to zero. The sequence $\left\{ x^{*}\right\}
=\left\{ x_{p}^{*}:p=2,4,8,...\right\} $ is an element of a $\ell _{\infty }$
Banach space (sup norm). The collection of fixed point equations 
\[
f_{\mu ^{*}}^{(p)}(x_{p}^{*},c)-x_{p}^{*}=0 
\]
defines a $C^{\infty }-$mapping $F(x^{*},c)$ from $\ell _{\infty }\times
R\rightarrow \ell _{\infty }$ . Because $\left( f^{(p)}\right) ^{^{\prime
}}(x^{*})$ at $c=0$ is negative and bounded for all $p$, the derivative $%
D_{1}F$ of the mapping in the first argument is invertible. Therefore, by
the implicit function theorem for Banach spaces, there is a $c^{*}$ such
that for $c<c^{*}$ the function $x^{*}(c):R\rightarrow \ell _{\infty }$ is
defined, that is, there are $p-$periodic orbits for all periods $p=2^{n}$.

For the uncoupled case ($c=0$) the instability factor $\left( f^{(p)}\right)
^{^{\prime }}(x^{*})$ for each mapping is the product $\left( -2\mu
_{*}\right) ^{p}\prod_{k=1}^{p}x(k)$ over the orbit coordinates. For $%
c<c^{*} $, the orbit structure being preserved, their projections on the
axis are continuous deformations of the $c=0$ case which will preserve the
geometric relations of the Feigenbaum accumulation point. Hence the same
products for the projected coordinates suffer changes of order $a(c)p\lambda
^{p}$ and remain bounded.

\textit{Second step}: Stabilization of at least one orbit for all periods $%
p=2^n $ with $n>N(c)$

The stability of the periodic orbits is controlled by the eigenvalues of the
Jacobian $J_{p}=\frac{Df_{\mu _{*},c}^{(p)}}{Dx}$ in the fixed point of $%
f_{\mu _{*},c}^{(p)}$. The map (\ref{5.1}) is a composition of two maps $%
f_{1}\circ f_{2}$%
\[
f_{1}:\left( 
\begin{array}{c}
x_{1} \\ 
x_{2}
\end{array}
\right) \rightarrow \left( 
\begin{array}{c}
1-\mu _{*}x_{1}^{2} \\ 
1-\mu _{*}x_{2}^{2}
\end{array}
\right) 
\]
\[
f_{2}:\left( 
\begin{array}{c}
x_{1} \\ 
x_{2}
\end{array}
\right) \rightarrow \left( 
\begin{array}{c}
(1-c)x_{1}+cx_{2} \\ 
cx_{1}+(1-c)x_{2}
\end{array}
\right) 
\]
and by the chain rule the Jacobian has determinant 
\begin{equation}
\det J_{p}=(1-2c)^{p}(-2\mu _{*})^{2p}\prod_{k=1}^{p}x_{1}(k)x_{2}(k)
\label{5.2}
\end{equation}
Because of the permanence of the periodic orbits, for small $c,$ in the
neighborhood of the original coordinates (those for $c=0$), the product of
the last two factors in (\ref{5.2}) is uniformly bounded for all $p$. Then
for all sufficiently large $p$, $\left| \det J_{p}\right| <1$. The question
is how this overall contraction is distributed among the two eigenvalues of $%
J_{p}$.

To discuss the nature of the eigenvalues we may use a first order
approximation in $c$. For a periodic orbit of period $p=2^{n}$ we define 
\begin{equation}
X_{l,q}^{(i)}=\left\{ 
\begin{array}{lc}
(-2\mu _{*})^{q-l+1}\prod_{k=l}^{q}x_{i}(k) & \textnormal{if }q\geq l \\ 
1 & \textnormal{if }l>q
\end{array}
\right\}  \label{5.3}
\end{equation}
For small $c$ consider the linear approximation to the Jacobian $J_{p}$%
\[
\left( 
\begin{array}{cc}
(1-pc)X_{1,p}^{(1)} & c\sum_{k=1}^{p}X_{1,k}^{(1)}X_{k+1,p}^{(2)} \\ 
c\sum_{k=1}^{p}X_{1,k}^{(2)}X_{k+1,p}^{(1)} & (1-pc)X_{1,p}^{(2)}
\end{array}
\right) 
\]
The eigenvalues are 
\begin{equation}
\begin{array}{ccl}
\lambda _{\pm } & = & \frac{1}{2}(1-pc)\left(
X_{1,p}^{(1)}+X_{1,p}^{(2)}\right) \\ 
&  & \pm \frac{1}{2}\sqrt{(1-pc)^{2}\left(
X_{1,p}^{(1)}-X_{1,p}^{(2)}\right)
^{2}+4c^{2}\sum_{k=1}^{p}X_{1,k}^{(1)}X_{k+1,p}^{(2)}\sum_{k^{^{\prime
}}=1}^{p}X_{1,k^{^{\prime }}}^{(2)}X_{k^{^{\prime }}+1,p}^{(1)}}
\end{array}
\label{5.4}
\end{equation}
If the periodic orbit runs with the two coordinates $x_{1}$ and $x_{2}$
synchronized then 
\[
\begin{array}{ccc}
\lambda _{+} & = & X_{1,p} \\ 
\lambda _{-} & = & \left( 1-2pc\right) X_{1,p}
\end{array}
\]
and the orbit being unstable for $c=0$ it remains unstable for $c\neq 0$.
However if the two coordinates are out of phase by $\frac{p}{2}$ steps the
radical in Eq.(\ref{5.4}) is 
\[
\sqrt{\prod_{k=1}^{p}x(k)}\left| \sum_{i=1}^{p}x(i)x(i+1)\cdots x(i+\frac{p}{%
2}-1)\right| 
\]
The existence of a superstable orbit for all periods $p=2^{n}$ implies that
at $\mu =\mu _{*}$ the product $\prod_{k=1}^{p}x(k)$ has an odd number of
negative-valued coordinates. Therefore the two eigenvalues are complex
conjugate and, for small $c$, the contraction implicit in (\ref{5.2}) is
equally distributed by the two eigenvalues. Therefore for sufficiently large 
$N$ all orbits of this type with $p>2^{N}$ become stable periodic orbits. \ 
$\Box$

The attracting periodic orbits of the coupled system being associated to the
unstable periodic orbits of the Feigenbaum cascade, the basins of attraction
will be controlled by the neighborhoods of these orbits, in each coordinate.
Therefore a checkerboard-type structure is expected for the basins of
attraction.

\end{document}